\documentclass{aa}
\usepackage{graphicx}
\usepackage{txfonts}
\usepackage{gensymb}
\usepackage{color}      
\begin{document}

   \title{An evolutionary model for the V404~Cyg system}
   \titlerunning{An evolutionary model for the V404~Cyg system}
   \authorrunning{Bartolomeo Koninckx et. al.}

   \author{L. Bartolomeo Koninckx$^{1,2}$\thanks{Fellow of the Consejo Nacional de Investigaciones Cient\'ificas y T\'ecnicas (CONICET), Argentina, E-mail: leandrobart96@fcaglp.unlp.edu.ar}, M. A. De Vito$^{1,2}$\thanks{Member of the Carrera del Investigador Cient\'ifico, CONICET, Argentina.}, O. G. Benvenuto$^{1,2}$\thanks{Member of the Carrera del Investigador Cient\'ifico, Comisi\'on de Investigaciones Cient\'ificas de la Provincia de Buenos Aires (CIC), Argentina.}}

   \institute{Instituto de Astrof\'isica de La Plata, IALP, CCT-CONICET-UNLP, Argentina,
         \and
             Facultad de Ciencias Astron\'omicas y Geof\'isicas de La Plata, Paseo del Bosque S/N, (1900) La Plata, Argentina.
             }

   \date{Received April 03, 2023; accepted April 23, 2023}

 
  \abstract
   {V404~Cyg is a low mass X-Ray binary (LMXB) system that has undergone outbursts in 1938, 1989, and 2015. During these events, it has been possible to make determinations for the relevant data of the system. This data include the mass of the compact object (i.e., a black hole; BH) and its companion, the orbital period, the companion spectral type, and luminosity class. Remarkably, the companion star has a metallicity value that is appreciably higher than solar. All these data allow for the construction of theoretical models to account for its structure, determine its initial configuration, and predict its fate. Assuming that the BH is already formed when the primary star reaches the zero age main sequence, we used our binary evolution code for this purpose. We find that the current characteristics of the system are nicely accounted for by a model with initial masses of 9~M$_\odot$ for the BH, 1.5~M$_\odot$ for the companion star and an initial orbital period of 1.5~d, while also considering that at most $30\%$ of the mass transferred by the donor is accreted by the BH. The metallicity of the donor for our best fit is $Z=0.028$ (twice solar metallicity). We also studied the evolution of the BH spin parameter, assuming that is not rotating initially. Remarkably, the spin of the BHs in our models is far from reaching the available observational determination. This may indicate that the BH in V404~Cyg was initially spinning, a result that may be relevant for understanding the formation BHs in the context of LMXB systems.}


   \maketitle
%

\section{Introduction}

Close binary systems with a black hole (BH) component have been studied since the first detection of accreting BHs in binary systems by Roche lobe overflow (RLOF)  in the 1960 decade, alongside with the first missions containing X-ray detectors (see, e.g., \citealt{1962PhRvL...9..439G,1967AJ.....72R.812L}). The material lost by a normal companion of low mass, known as low mass X-Ray binary (LMXB) or of high mass, namely, a high mass X-Ray binary (HMXB) forms an accretion disk around the BH. Mass and angular momentum are thereby transferred to the BH, releasing an intense X-ray flux.  This particular group of binary systems has been studied from both, an observational and a theoretical point of view (among the most recent works: e.g., \citealt{2023ApJ...945...65Y}, \citealt{2022ApJ...930....9M},
\citet{2021MNRAS.506..581M}, \citealt{2021ApJ...912...86F}, \citealt{2020A&A...638A..39L}, \citealt{2017ApJ...843L..30I})

V404~Cyg is a member of the LMXB family. It was discovered by the space satellite \textit{Ginga} in May of 1989 as the transient X-ray source GS 2023+338 \citep{Makino89}. Its optic counterpart was identified as the variable star V404~Cyg \citep{Wagner89}. 
Later, \citet{1989ESASP.296..103C} identified the source as an LMXB. The binary has an orbital period of $P = 6.473\pm 0.001$~d \citep{Casares92} and a mass function of $f(M) = 6.08\pm0.06$ M$_\odot$ \citep{Casares94}. This high value of the mass function suggests the nature of the accretor is that of a BH. The mass ratio was determined by \citet{Casares92} as $q\equiv{M_{\rm d}}/{M_{\rm BH}}=0.06^{+0.004}_{-0.005}$, with $M_{\rm d}$ and $M_{\rm BH}$ as the masses of the donor star and the BH, respectively. The companion was confirmed as a giant star when \citet{Khargharia2010} determined its spectral type as K3~III. They also found the binary's inclination, $i= 67$\degree$^{+3\degree}_{\rm -1\degree}$. Knowing all these parameters, the determination of the masses of the components is immediate, namely, $M_{\rm d} = 0.54\pm0.05$ M$_\odot$ and $M_{\rm BH} = 9.0^{+0.2}_{\rm -0.6}$ M$_\odot$, for the donor and the BH, respectively. 
In 2009 a precise estimation of the distance was taken, giving a value of $d=2.39\pm0.14$~kpc by the measure of the parallax of the system on radio waves \citep{Miller-Jones2009}. \citet{Ziolkowski2018} (henceforth ZZ18) presented, based on these observational data, the  radius  of the donor star, $R_{\rm d} = 5.50^{+0.17}_{-0.18}$ R$_\odot$, the effective temperature (from the spectral type) $T_{\rm eff}=4274^{+116}_{-113}$~K and the luminosity of the donor star, $L_{\rm d}=8.7^{+1.7}_{-1.4}$ L$_\odot$.\\
Many studies have sought to calculate the accretion rate onto the compact object during the various outburst that had taken place in 1938, 1989, and 2015 \citep{Chen97,Zycki99,Motta2017a}. The system presented two episodes of outburst in 2015, namely: in July \citep{Barthelmy2015} and in December \citep{Marti2016,Motta2016}. Due to the large absorption reported \citep{Kimura2016a}, it was challenging to construct an X-ray luminosity curve this year and, thus, to estimate a mass accreted during this event. With the information provided by the above-mentioned authors, ZZ18 stated that the value $\langle \dot{M}_{\rm BH} \rangle = 4.0 \times 10^{-10}$~M$_\odot$ \,yr$^{-1}$ is most likely an upper limit for the accretion rate onto the BH in V404~Cyg. This value is  lower than the estimated mass loss rate for the donor star $\langle -\dot{M_{\rm d}} \rangle =$ $1.1$ $\times$ $10^{-9}$ M$_\odot$ \,yr$^{-1}$ that was predicted using the equation 25a from \citealt{Webbink83} for the system V404~Cyg. ZZ18 also re-obtained these values using an updated evolutionary model. The difference between the amount of mass lost by the donor and the mass accreted by the BH has been attributed to the mass that gets lost from the system, advecting angular momentum along with it. These mass and angular momentum losses make the system evolve in a non-conservative way. In addition, there are observational indications that V404~Cyg is currently losing mass \citep{2016Natur.534...75M}.

In this work, we consider non-conservative close binary evolutionary models with the objective of reproducing observational data available for the main parameters of V404~Cyg, with the aim of obtaining a possible progenitor for the system. We also predict possible results for the evolution of the system's donor and show some theoretical parameters that we expect to be observationally measured in the future, such as the time derivative of the orbital period. On the other hand, we also study the evolution of the spin parameter in the context of our models. We specify the numerical code used in Section~\ref{sec:code}, show the results of our models in Section~\ref{sec:results}, and present our conclusions in Section~\ref{sec:conclusions}.


\section{The binary evolution code} \label{sec:code}
The main tool for this work is the binary evolutionary code described in \citet{Benvenuto2003,DeVito2012,BDVHa}.
When components remain detached, it works as a standard evolutionary code for isolated stars. In the case of semi-detached configurations, our code includes the mass transfer rate, $\dot{M}_1$, as a new variable in the difference equations. Then, the mass of the donor is $M_1= M^{1, prev} + \dot{M}_1 \Delta t$, where $M_{1, prev}$ is the mass of the donor in the previous stage and $\Delta t$ is the time step. As $M_1$ appears in the equations of the structure of the entire model,  $\dot{M}_1$ is treated as a {\it global} variable to be solved. This is in contrast with all the other variables that are local and meant to be relaxed. When handling the corresponding generalized Henyey matrix, this treatment involves a non-zero column. The resulting matrix equation can be solved with a slight modification of the standard algebra. This solves the structure of the donor star, the orbital evolution, and the value of the mass transfer rate simultaneously in a fully implicit way, which makes the algorithm numerically stable. A detailed explanation of the procedure is given in \cite{Benvenuto2003}. We assume that the mass is only transferred via Roche lobe overflow (RLOF). As for opacities, we used OPAL libraries \citep{Iglesias96} for temperatures of $T\geq 10^4$~K and molecular opacities computed by \citet{Feguson2005} for lower values of $T$. A detailed description of how the code works may be found in \citet{BDVHa}. 

Regarding the abundances assumed for our models, in the first step of our calculations, we followed ZZ18 to employ the solar metallicities. We have set them to $X=0.710$, $Y=0.276$, and $Z=0.014$, whereas the mixing length parameter has been set to $\alpha_{\rm MLT}=1.50$. With these values, our code is able to compute a solar structure compatible with observations at its present age. We remark here that these abundances are slightly different from those given in \citet{2021A&A...653A.141A} who measured values of $X= 0.7438 \pm 0.0054$, $Y= 0.2423 \pm 0.0054$, and $Z= 0.0139 \pm 0.0006$ at the surface of the Sun. If we set the abundances to these values, the Sun would be slightly under-luminous by 0.05 dex, which is a small discrepancy since the physical ingredients employed by \citet{2021A&A...653A.141A} are different from ours. So, we decided to slightly adjust the initial abundances to produce a Sun compatible with observations. In the second step of the model calculations, we took into account the determination of abundances for the donor star in V404~Cyg, presented in \citet{2011ApJ...738...95G}, and we employed  $X=0.71$, $Y=0.262$ and $Z=0.028$.
\subsection{Non-conservative mass transfer and orbital evolution}\label{Nonconservative}
For cases of conservative binary evolution calculations, total mass, and orbital angular momentum remain as constants. However, in analyzing the difference between the estimated mass loss rate from the donor component and the estimated accretion rate on the BH from V404~Cyg, it is commonly assumed that mass gets lost in a non-conservative mass transfer episode advecting angular momentum away from the system (\citealt{Webbink83,Chen97,Zycki99,Motta2017a}; ZZ18). This is expected to occur in some astrophysical scenarios of interest, so this phenomenon was included in the calculations.\\
We employed the usual equation to compute the evolution of the orbital semi-axis. This can be obtained using the definition of the angular moment combined with Kepler's Third Law. The episode of non-conservative mass transfer is specified by two free parameters, as in \citet{Rappaport82,Rappaport83}: 1) the fraction $\beta$ of mass lost by the primary star\footnote{We will name as primary (i.e., with the sub-index 1) to the object that starts losing mass. In this case, we will name this way to the donor star and we will refer to the BH as secondary with the sub-index 2.} that is accreted by the secondary star, and 2) the specific angular momentum of matter lost away from the system $\alpha$ in units of the same quantity for the compact object.
We assume that the orbit is always well approximated by a circle of radius $r_{\rm orb}$ (where $r_{\rm orb}$ is a function of time) and we have neglected the rotational angular momentum of the components. In the case where the angular momentum is lost only by mass ejection from the system,\typeout{Equation~(\ref{dJ}) can be rewritten using Kepler's third law and the expression of total angular as}\typeout{, as the following differential equation} we find:
\begin{equation}\label{MassEjection}
    \frac{dJ_{\rm ME}}{dt} = \alpha (1-\beta)\sqrt{GMr_{\rm orb}}\; \bigg(\frac{M_{\rm 1}}{M}\bigg)^2\; \dot{M_{\rm 1}},
\end{equation}where $G$ is the gravitational constant and $M=M_1+M_2$ is the total mass of the system, with $M_{\rm 1}$ and $M_{\rm 2}$ the masses for the donor star and the BH, respectively.\\
Angular momentum can also be lost from the system by gravitational radiation and it is calculated according to the standard formula \citep{Landau71}:
\begin{equation}\label{GravitationalRadiation}
    \frac{d ln(J_{\rm RG})}{dt}=-\frac{32G^3\mu}{5c^5}\frac{M^2}{r_{\rm orb}^4},
\end{equation} where $c$ is the vacuum speed of light and $\mu=\frac{M_{\rm 1} M_{\rm 2}}{M}$.\\
The code also considers angular momentum loss due to magnetic braking, using the prescription of \citet{Rappaport83}, based on the magnetic-braking law of \citet{Verbunt81},
\begin{equation}\label{Magneticbraking}
    \frac{dJ_{\rm MB}}{dt}=-3.8\times10^{-30} M_{\rm 1} R_{\rm 1}^4 \omega^3 \,\mathrm{dyn} \,\mathrm{cm},
\end{equation} where $\omega$ is the angular rotation frequency of the donor star, assumed to be synchronized with the orbit, and $R_{\rm 1}$ is the donor's radius. The code includes full magnetic braking when the star has a sizable convective envelope embracing a mass fraction $\geq$ $0.02$.\\ 
Replacing Equations~(\ref{MassEjection}) - (\ref{Magneticbraking}) in the expression for the evolution of the angular momentum and considering $\dot{M}_{\rm 2}=-\beta \dot{M}_{\rm 1}$ from the definition of $\beta$, we obtain a differential equation for the orbital separation, which has no analytical solution.\\
\subsection{Eddington limit and black hole spin parameter} 

This is the first work in which we employ our code to calculate the evolution of a binary system with a BH; in previous papers, the companion was a neutron star or a normal star. Thus, we had to change the accretion efficiency of the compact object. Furthermore, we are interested in calculating the evolution of the spin angular momentum of the BH as it receives mass and angular momentum from its companion. For that purpose, we followed the prescriptions given in \citet{Podsi2003} (henceforth, PRH03):\\
The luminosity released due to accretion onto the BH is: 
\begin{equation}\label{eq:bhluminosity}
    L_{\rm 2}=\eta \dot{M}_{\rm 2}c^2,     
\end{equation}    
where  $\dot{M}_2$ is the BH accretion rate, $c$ is the vacuum speed of light, and $\eta$ is the efficiency with which the BH radiates, determined by the last stable particle orbit. This parameter can be expressed as:
\begin{equation}\label{eq:eficiencia}    
\eta=1-\sqrt{1-\left( \frac{M_{\rm 2}}{3M^0_{\rm BH}} \right)},
\end{equation}where the quantities $M^0_{\rm BH}$ and $M_{\rm 2}$ are the initial and present mass of the BH, respectively.

Equaling the BH luminosity $L_2$ with Eddington's luminosity and assuming spherical accretion, an expression for the maximum accretion rate onto de BH can be obtained (see PRH03) as:
\begin{equation}
    \dot{M}_{\rm Edd} \simeq 2.6 \times 10^{-7} \mathrm{M}_{\odot} \mathrm{yr}^{-1} \left( \frac{M_{\rm 2}}{10 \mathrm{ M}_{\odot}} \right) \left( \frac{\eta}{0.1} \right)^{-1} \left( \frac{1+X}{1.7} \right)^{-1},
\end{equation} where $X$ is the hydrogen mass fraction. The BH accretion rate is limited by this value through its evolution. \\
On the other hand, the accretion phenomena not only affects the mass of the BH. As it accretes matter, and since that matter carries angular momentum with it, the spin parameter of the BH defined as $a^*\equiv cJ_2/GM_2^2$ also increases, according to:
\begin{equation}\label{BHspinparameter}
    a^*=\left( \frac{2}{3} \right)^{\frac{1}{2}} \frac{M^0_{\rm BH}}{M_{\rm 2}} \left\{ 4-\left[ 18\left( \frac{M^0_{\rm BH}}{M_{\rm 2}} \right)^2 -2\right]^{\frac{1}{2}} \right\},
\end{equation}
It is important to remark that these expressions are adequate for an initially non-rotating BH (see PRZ03) and are valid when $M_{\rm 2} < \sqrt{6}M^0_{\rm BH}$, which is the case for all our calculations (for a detailed treatment see \citealt{Bardeen70,King99}).\\
\section{Models and results}\label{sec:results}
Our primary objective in this work is to obtain possible progenitors for the binary system V404~Cyg. With this goal, we analyzed the results generated by different sets of initial parameters: masses of the donor star and the BH ($M^0_{\rm d}$ and $M^0_{\rm BH}$, respectively), the orbital period $P^0_{\rm orb}$ and the fraction $\beta$ of the mass lost by the donor that is accreted by the BH (as defined in Section~\ref{Nonconservative}). We fixed the free parameter that describes the specific angular momentum of matter lost as $\alpha=1$.\\
An important consideration that stands out is that this code calculates the evolution of the donor star with an already existing BH. That is to say, we are not discussing how the compact object formed and we are also avoiding the common envelope phase. The code assumes that the orbit of the system components is circularized, so the circular restricted three-body problem can be applied. This is a very reasonable assumption for V404~Cyg, where we can get the orbital eccentricity using the definition of the mass function and the observational estimations for the orbital period, the K semi-amplitude, and the mass function given by \cite{Casares92} obtaining a value of $e \sim 0.024$.\\
\subsection{Models with solar abundances}

We computed 72 evolutionary sequences with solar abundances exploring the combinations of initial donor masses of 1.5 and 2.0~M$_\odot$, initial BH masses of 7, 8, and 9~M$_\odot$, initial orbital period of 0.75, 1.00, and 1.25~days and values of 0.1, 0.3, 0.7 and 1.0 for the $\beta$ fraction. The identification of each model was built using the first character associated with the initial masses of the model (see Table~\ref{table:model_names}), followed by labels that are related to the initial orbital period and the $\beta$ parameter. For example, the model C\_075\_07 was calculated with initial masses of 1.5 M$_\odot$ and 8 M$_\odot$ for the donor and the BH, respectively, as well as an initial orbital period of 0.75 d and $\beta = 0.7$.

\begin{table}
    \centering
    \caption{Groups of models divided by the initial masses considered for the donor (1.5 and 2.0 M$_\odot$) and for the BH (7, 8, and 9~M$_\odot$).}\label{tab:group}
    \begin{tabular}{c|c|c}
    \hline\hline
        \textbf{Group} & \textbf{M}$\mathbf{^0_{\rm d}}$\textbf{[M}$_\odot$\textbf{]}& \textbf{M}$\mathbf{^0_{\rm BH}}$\textbf{[M}$_\odot$\textbf{]}\\
        \hline
        A & 1.5 & 7\\
        B & 2.0 & 7\\
        C & 1.5 & 8\\
        D & 2.0 & 8\\
        E & 1.5 & 9\\
        F & 2.0 & 9\\
        \hline
    \end{tabular}
    \label{table:model_names}
\end{table}
We introduce the function $\epsilon^2$ that helps us to determine how close the parameters given by our models are from the ones observationally acquired. This quantity is defined as:
\begin{equation}
    \begin{aligned}
        \epsilon^2=\sum_i\epsilon_i^2\hspace{1mm},\hspace{4mm} 
        \epsilon_i=\frac{E_{\rm i}-E^{\rm obs}_{\rm i}}{E^{\rm obs}_{\rm i}}
    \end{aligned}
    \label{eq:epsilon}
\end{equation}where $E_{\rm i}$ and $E^{\rm obs}_i$ are the model and observational data for the parameter $i$, and $i=1,...,5$ correspond to the BH mass ($M_{\rm BH}$), the donor's mass ($M_{\rm d}$), the orbital period ($P_{\rm orb}$), effective temperature ($T_{\rm eff}$) and luminosity ($L_{\rm d}$)\typeout{ or radius ($R_{\rm d}$)} of the donor star, respectively. For example, $\epsilon_{\rm 1}= (E_{\rm 1}-E^{\rm obs}_1)/E^{\rm obs}_1$ means $\epsilon_{M_{\rm BH}}=(M_{\rm 2}-M_{\rm BH})/M_{\rm BH}$. The time dependence for $\epsilon_i^2$ will be given by the change of the quantities along the evolutionary sequences.\typeout{on the modeled parameter on the evolution code.} All values used for the observational measures can be found in Table~\ref{tab:obsest}. For $\epsilon^2=0$, we have a case where all the $i$ parameters calculated from the models are equal to the ones obtained observationally. This makes it easier to see that this quantity helps us to determine when and how the models represent the system's observational data simultaneously.\\
\begin{figure}
    \centering
    \includegraphics[width=\columnwidth]{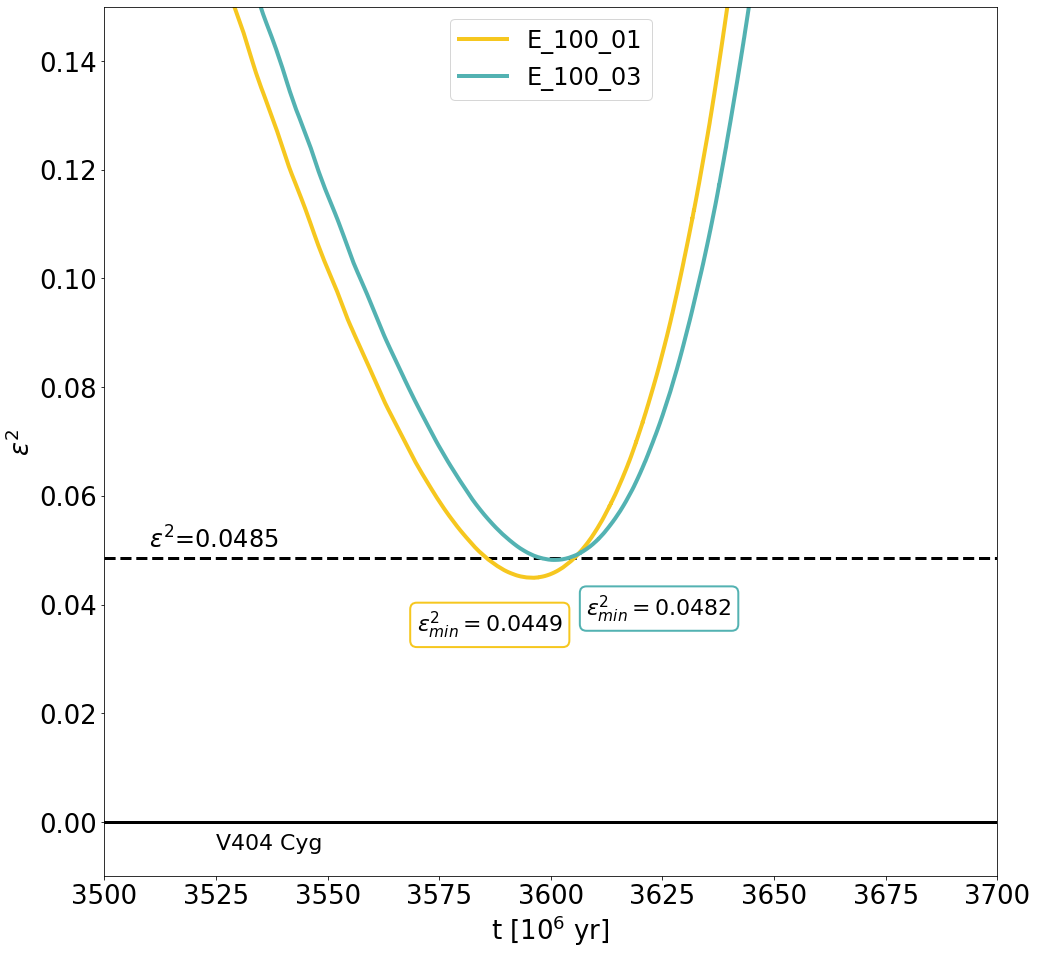}
    \caption{Quantity $\epsilon^2$ as a function of time for our best models with solar composition: E\_100\_01 and E\_100\_03. With a dashed line and a grey area is indicated the value of $\epsilon^2 = 0.0485$ and the acceptance region we considered for our models. With a black solid line, the value $\epsilon^2 = 0$ represents the situation where all the parameters modeled are equal to the ones observed simultaneously.
    }
    \label{fig:epsi}
\end{figure}
Computing this function along the calculated sequences allows us to obtain a minimum value for each of them.
This minimum value corresponds to the time when the quantities modeled are closest to the ones observationally estimated. We consider that the model represents the characteristics of the observed system well enough when the minimum value of the $\epsilon^2$ function is lower than $0.0485$, which is the value of the sum obtained when $E_{\rm i} = E_i^{\rm obs} + \sigma_{\rm i}^{\rm obs}$ for each $i$, where $\sigma_{\rm i}^{\rm obs}$ is the observational error for each parameter that can be found on Table~\ref{tab:obsest}. This restriction guarantees that the modeled parameters are not far from their observational uncertainties. Only two of our models calculated with solar abundances satisfy this condition, namely: E\_100\_01 with $\epsilon_{min}^2 = 0.0449$ and E\_100\_03 with $\epsilon_{min}^2 = 0.0482$, as seen on Figure~\ref{fig:epsi}. In Sections~\ref{sec:masstransfer} and~\ref{sec:periodBHspin}, we present our results for these models.\\
\begin{table}
    \centering
    \caption{Observational data for V404~Cyg system. Each quantity comes accompanied with the corresponding error (Column 3) and the reference where it was taken from (Column 4).}
    \label{tab:obsest}
    \begin{tabular}{p{0.15\columnwidth} p{0.12\columnwidth} p{0.1\columnwidth} p{0.45\columnwidth}}
    \hline\hline
        \textbf{Parameter} & \textbf{Value} & \textbf{Error} &     \textbf{References} \\
        \hline
        $M_{\rm BH}$ & $9$ M$_\odot$ & $^{+0.2}_{-0.6}$ &     \citet{Casares94,Khargharia2010} \\
        \hline
        $M_{\rm d}$ & $0.54$ M$_\odot$ & $\pm 0.05 $ &     \citet{Casares94,Khargharia2010} \\
        \hline
        $P_{\rm orb}$ & $6.47$ d & $\pm 0.001$ & \citet{1989ESASP.296..103C} \\
        \hline
        $T_{\rm eff}$ & $4274$~K & $^{+116}_{-113}$ & \citet{Cox2000,Khargharia2010} \\
        \hline
        $L_{\rm d}$ & $8.7$ L$_\odot$ & $^{+1.7}_{-1.4}$ & \citet{Cox2000,Khargharia2010}; ZZ18 \\
        \hline
    \end{tabular}
\end{table}

\subsubsection{Mass transfer}\label{sec:masstransfer}
\begin{figure}
    \centering
    \includegraphics[width=\columnwidth]{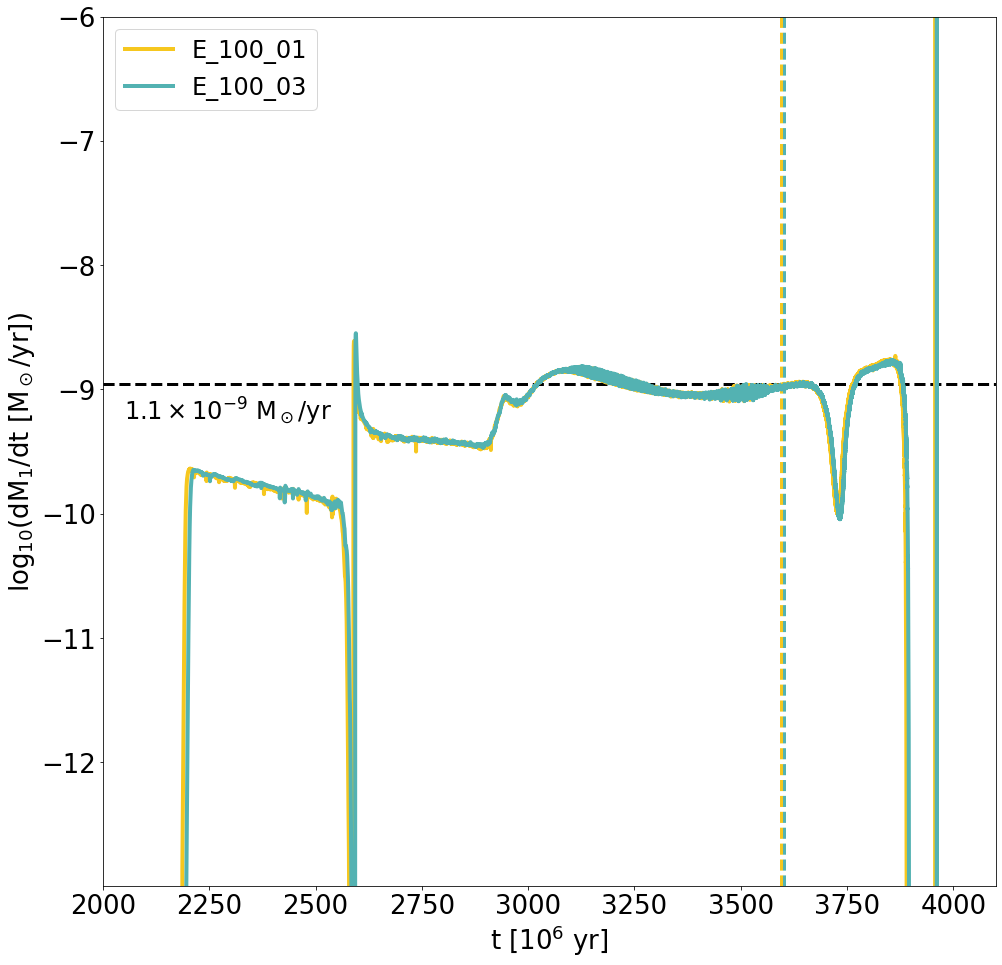}
    \caption{Donor star mass loss rate for each of the best models with solar composition: E\_100\_01 and E\_100\_03. As the only variation between these models is on the $\beta$ value, and this parameter does not affect strongly the mass loss episode, their plots are mostly overlapping. The black dashed horizontal lines represent the estimation for the mass loss rate for V404Cyg of $\dot{M}_{\rm d}=1.1\times10^{-9}$ M$_\odot$ yr$^{-1}$ is denoted with a dashed horizontal line. As for the dashed vertical lines, they represent the times of the minimum value of the epsilon squared function.}
    \label{fig:LossRate}
\end{figure}

\begin{figure}
    \centering
    \includegraphics[width=\columnwidth]{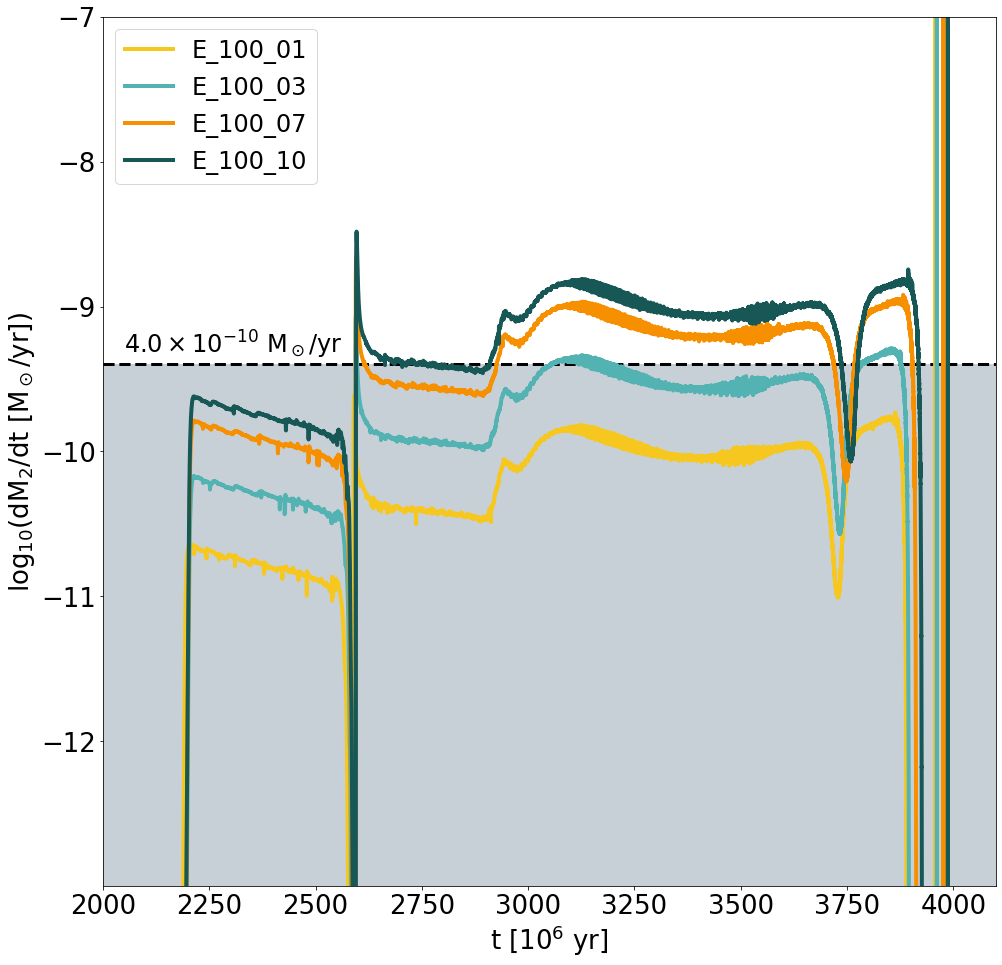}
    \caption{BH accretion rate for the models with solar abundances of the E group with P$_{\rm orb}^0 = 1$~d. The qualitative form of the functions is related to the mass loss rate of the donor star by the relation $\dot{M}_{\rm 2}=-\beta\dot{M}_{\rm 1}$. This is, for a higher value for $\beta$ larger would be the accretion rate onto the compact object. The shaded area represents the zone below the likely upper limit for the mass accretion rate $\left(\left<\dot{M}_{\rm 2}\right>=4\times10^{-10} \rm{M}_\odot \rm{yr}^{-1}\right)$ given by ZZ18. Models with $\beta\leq0.3$ agree with this upper limit and the results reached by the mentioned authors.}
    \label{fig:AcRate}
\end{figure}
\typeout{As we already said, the observational data for the mass transfer episode for V404~Cyg seems to imply that a part of the mass gets lost from the system, advecting angular momentum away.\\}
Among other studies, ZZ18 have studied the non-conservative mass transfer episode for this system. In their work, they estimated the accretion rate over intervals between outbursts and got a likely upper limit for it, with a value of $\langle\dot{M}_{\rm BH}\rangle=4\times10^{-10}$ M$_\odot\; \rm{yr}^{-1}$. They also stated that their models with $\beta \lesssim 0.33$ were the ones that are in agreement with this limit. For the donor mass loss rate, the existing estimation is of $\dot{M}_{\rm d}=1.1\times10^{-9}$ M$_\odot\; \rm{yr}^{-1}$, value taken from \citet{Webbink83} using the equation 25a with the V404~Cyg system's parameters.\\

Our results for the mass transfer episode are resumed in Figure~\ref{fig:LossRate}, where the donor mass loss rate ($\dot{M}_{\rm 1}$) for the best models is shown. In Figure~\ref{fig:AcRate}, we show the accretion rate on the compact object ($\dot{M}_{\rm 2}$) for the same models.\\
For the first case (shown in Figure~\ref{fig:LossRate}), the existing estimation has been featured with a horizontal dashed line. We have highlighted the age predicted by our models when the epsilon squared function reaches its minimum value (see Figure~\ref{fig:epsi}) with a vertical dashed line. At this time, the mass loss rate of the donor star for models E\_100\_01 and E\_100\_03 nicely agrees with the above-quoted estimation\typeout{ while for the conservative model C\_125\_10 we find an appreciable difference (of $\sim37 \%$)}.\\
As for the mass accretion rate onto the BH (Figure~\ref{fig:AcRate}), we added to our two best models the ones from the same group and initial orbital period so the effect of the variation of $\beta$ becomes evident. As is typical in the computation of close binary models, this parameter relates the mass loss by the donor with the one that is accreted by the compact object as $\dot{M}_{\rm 2}=-\beta\dot{M}_{\rm 1}$. This relation makes the mass accretion rate function similar in form to the mass loss rate function of the previous figure, but a change in the $\beta$ parameter does not provoke a strong variation to $\dot{M}_{\rm 1}$ as to $\dot{M}_{\rm 2}$. The shaded area on this figure represents the values below the likely upper limit suggested by ZZ18. We found that both of our models with $\beta\leq 0.3$ remains under this limit, confirming previous results.
\subsubsection{Orbital period and BH spin parameter}\label{sec:periodBHspin}
\begin{figure}
    \centering
    \includegraphics[width=\columnwidth]{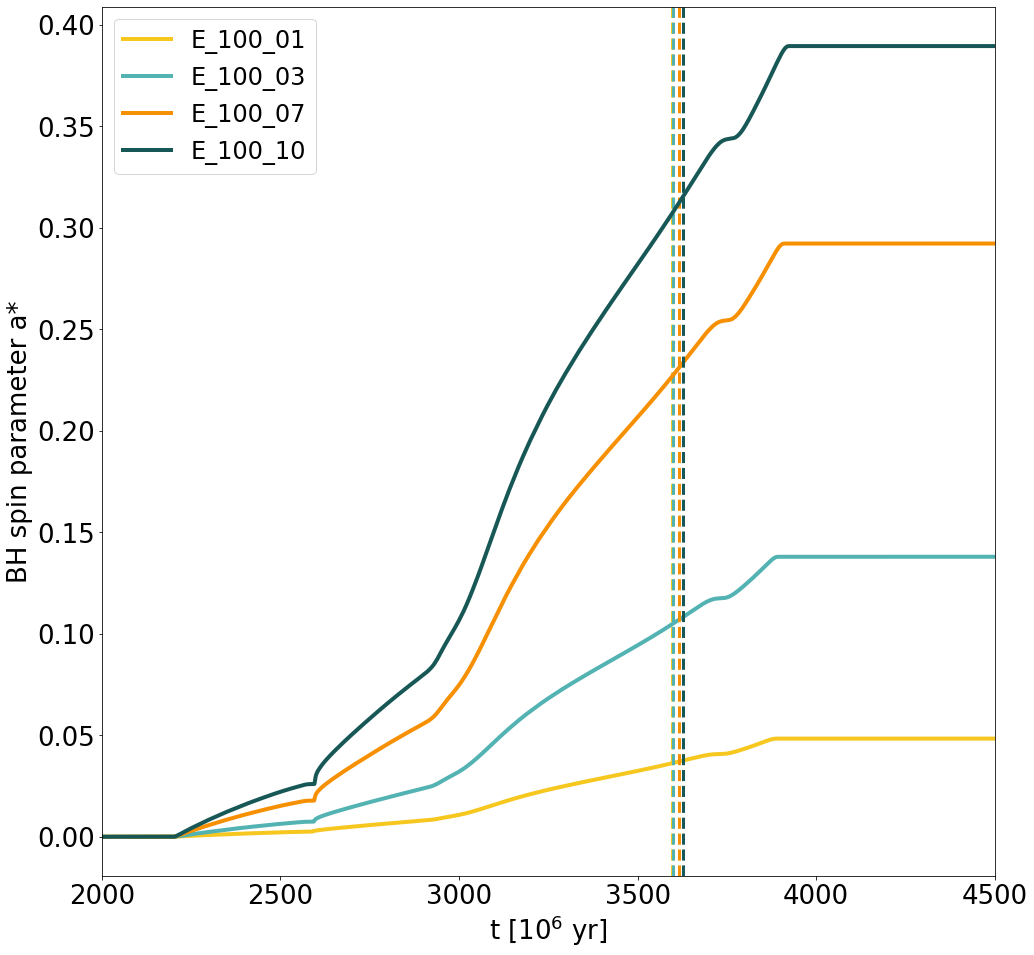}
    \caption{BH spin parameter for the V404~Cyg compact object for each of the best models, according to Equation~(\ref{BHspinparameter}). As the BH accretes matter it spins up, provoking the increase of $a^{*}=cJ_2/GM_2^2$. Thus, the lower the $\beta$, the lower the BH spin. The vertical dashed lines are the predicted ages for V404~Cyg. The values of the BH spin for these times are given in Table~\ref{tab:BHparam}.}
    \label{fig:BHspinparam}
\end{figure}
\begin{figure*}
    \centering
    \includegraphics[width=0.73\textwidth]{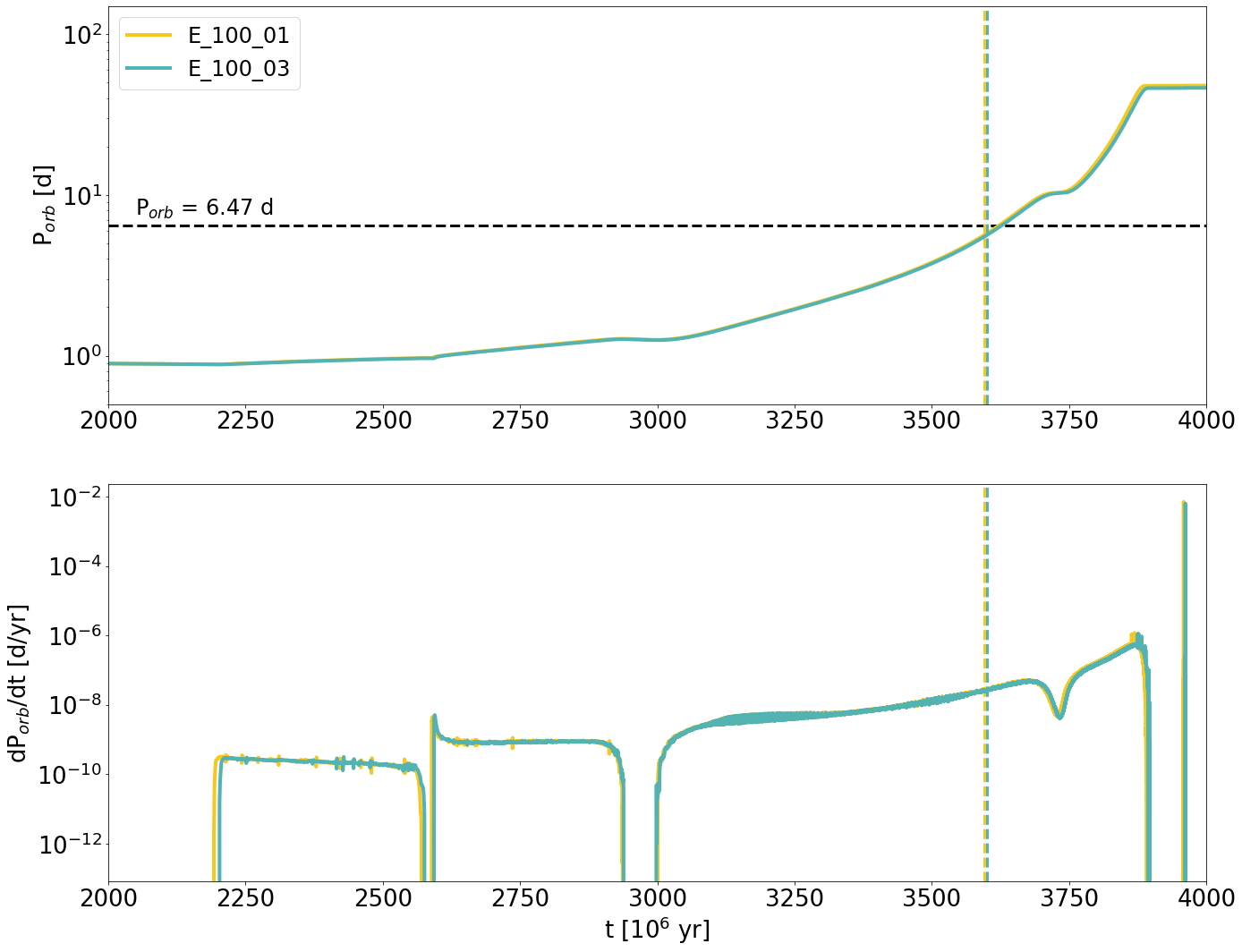}
    \caption{Orbital period as a function of time for our best models with solar metallicity (solid lines), observational estimation for the present orbital period $P_{\rm orb}=6.47$ d (dashed horizontal line) and age of the system for the system predicted for each model (dashed vertical lines) shown at the top. The bottom panel shows the time derivative for the orbital period for the same models and ages.}
    \label{fig:DPDT}
\end{figure*}
A non-conservative mass transfer episode in a binary system essentially consists of a fraction of the mass transferred by the donor being accreted by the companion, while the rest is lost from the system. The mass accreted by the BH accelerates its rotation. This phenomenon can be studied by analyzing the evolution of the BH spin parameter $a^{*}=cJ_{BH}/GM_{BH}^2$, with $J_{BH}$ as the rotational angular momentum of the BH.\\
The only available data for the spin parameter of V404~Cyg BHs is given in \citet{Walton2017}. This study used a spectral analysis of $NuSTAR$ X-ray observations of V404~Cyg for its 2015 outburst and proposed different models that fit the observations using the reflection method. Although these authors obtained multiple solutions for the BH spin parameter, they stated that the most robust one to be $a^{*}> 0.92$ with a $99\%$ of statistical uncertainty.\\
For the computation of this parameter, assuming that the BH is initially not rotating, we employed Equation \ref{BHspinparameter}. We obtained the evolution for the BH spin parameter over time, as shown in Figure~\ref{fig:BHspinparam}. In this graph (and Table~\ref{tab:BHparam}), we show our results for all the models from group E with an initial orbital period of 1~d, so it becomes evident that the larger the $\beta$ value, the faster the BH's final rotation.\\
\begin{table}
    \centering
    \caption{Value for the BH spin parameter $a^{*}$ at the time of the minimum on $\epsilon^2$ quantity. It becomes evident that the more mass the BH accretes, the more it finally spins up.}
    \label{tab:BHparam}
    \begin{tabular}{c|c|c}
        \hline\hline
        \textbf{Model} & \textbf{t}$\mathbf{_{\rm min}}$ \textbf{[Gyr]} & \textbf{BH spin parameter} $\boldsymbol{a}^{*}$ \\
        \hline
        E\_100\_01 & 3.596 & 0.04 \\
        E\_100\_03 & 3.601 & 0.11 \\
        E\_100\_07 & 3.616 & 0.23 \\
        E\_100\_10 & 3.628 & 0.32 \\
        \hline
    \end{tabular}
\end{table}
The solutions for the BH spin parameter at the presumed ages for the system (see Table~\ref{tab:BHparam}) correspond to a slow rotation regime when the existing estimations (even the ones for slow rotators obtained by \citealt{Walton2017}) are much higher.\\

\begin{table}
    \centering
    \caption{Orbital period with its time derivative value and the characteristic increase time-scale evaluated on the estimated age for the system for models with Z=0.014.}
    \label{tab:dpdt}
    \resizebox{\columnwidth}{!}{
    \begin{tabular}{c|c|c|c|c}
        \hline\hline
        \textbf{Model} & \textbf{t}$\mathbf{_{\rm min}}$ \textbf{[Gyr]} & \textbf{P}$\mathbf{_{\rm orb}}$\textbf{[d]}& $\dot{\mathbf{\rm P}}_{\mathbf{\rm orb}}$& \textbf{P}$\mathbf{_{\rm orb}}$\textbf{/}$\dot{\mathbf{\rm P}}_{\mathbf{\rm orb}}$ \textbf{[yr]}\\
        \hline
        E\_100\_01 & 3.596 & 5.65 & $7.49\times10^{-11}$ & $2.1\times10^{8}$ \\
        E\_100\_03 & 3.601 & 5.78 & $7.63\times10^{-11}$ & $2.0\times10^{8}$ \\
        \hline
    \end{tabular}
    }
\end{table}
As matter leaves the system it carries away angular momentum, as described by Equation~\ref{MassEjection}. Other effects that also modify the orbital period are gravitational radiation (Equation~\ref{GravitationalRadiation}) and magnetic braking (Equation~\ref{Magneticbraking}). In this work, we consider all three effects together, ultimately finding that the orbital period mostly increases with time, reaching values of P$_{\rm orb}=$ 46 - 48 d at the end of our calculations (age of the donor star of $14$~Gyr). The evolution of this quantity for each of our best models is shown on the top panel of Figure~\ref{fig:DPDT}, where the initial orbital period is $P^0_{\rm orb}=1.00$ d for both. Our results are in good agreement with the well-determined value of $P_{\rm orb}=6.47$~d, when $\epsilon^2$ has its minimum value. As for the bottom panel, we show the time derivative of the orbital period of our models, calculated for each time as an approximation of an incremental quotient. The results for this quantity and the characteristic timescale ($P_{\rm orb}/\dot{P}_{\rm orb}$) at the expected ages of the system can be found in Table~\ref{tab:dpdt}. We found an increased timescale of $\sim 2\times10^8$ yr, which is in good agreement with the one predicted by ZZ18. 

Although the time derivative of the orbital period would be very useful to test the evolutionary scenario, this quantity is not yet known and there are no prospects for deriving it any time soon. For measuring such quantity with an adequate degree of certainty, we would need a time basis that is far longer than what is presently available. \citet{2021arXiv211203779K} stated that this time basis should be long enough for the radii of the donor star and its Roche lobe to vary at least on a density scale height, $H_{\rho}$ ($H_{\rho}\equiv -dr/d\ln{\rho}$), which corresponds to thousands of years. Any measurement in the near future will surely reflect the occurrence of short-timescale phenomena, neglected in our calculations. In this sense, the values of $\dot{P}_{\rm orb}$ we have presented above are related to the ingredients considered for a modeling of the evolution of the whole system (magnetic braking, gravitational radiation, and mass loss from the system).

\subsection{Models with higher metallicity}
\label{sec:otroZ}

\typeout{As V404~Cyg is an object that belongs to the field of our Galaxy, we begin our exploration using solar metallicity.} As described above and as done in ZZ18, we initially considered solar abundances. Nevertheless, \cite{2011ApJ...738...95G} presented a chemical abundance analysis for the donor star, and obtained $\rm{[Fe/H] = 0.23 \pm 0.19}$. This value is well addressed with a metallicity of $\rm{Z = 0.028}$, two times the value corresponding to the Sun. Therefore, we calculated 30 additional models with this new value of Z, fixing the hydrogen abundance on X = 0.71. For this instance, we fixed the value of the initial donor's mass at $1.5$ M$_\odot$ and explored only the values of $\beta = 0.3$ and $0.1$, based on the results obtained from the solar metallicity analysis. For the values of the initial BH mass, we still considered $\rm{M_{BH}^0 =}$8, 9, and 10~M$_\odot$, and we explored the same interval of initial orbital periods, $\rm{P_{orb}^0 =}$0.75, 1.00, and 1.25 d, adding 1.50 and 1.75 d values to the analysis. These models are identified similarly to those corresponding to solar metallicity, but adding "$\_Z028$" at the end of the name.

\begin{figure}
    \centering
    \includegraphics[width=\columnwidth]{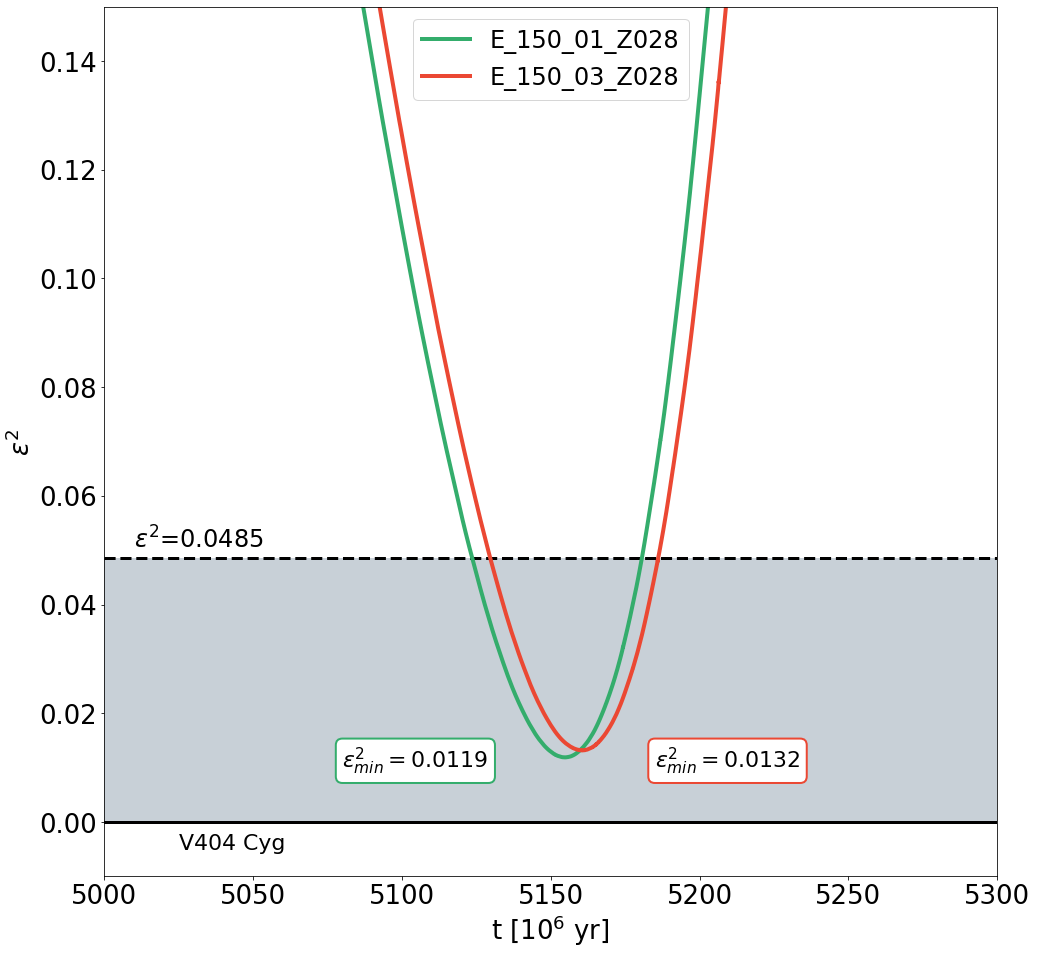}
    \caption{Quantity $\epsilon^2$ as a function of time for our best models with the change on metallicity: E\_150\_01\_Z028 and E\_150\_03\_Z028. The elements represented are the same as in Figure \ref{fig:epsi}. \typeout{With a dashed line and a grey area is indicated the value of $\epsilon^2 = 0.0485$ and the acceptance region we considered for our models. With a black solid line, the value $\epsilon^2 = 0$ represents the situation where all the parameters modeled are equal to the ones observed simultaneously.
    }}
    \label{fig:epsi_Z028}
\end{figure}
The best models we obtained are: E\_150\_01\_Z028 and E\_150\_03\_Z028, with minimum $\epsilon^2$ values of 0.0119 and 0.0132, respectively (see Figure~\ref{fig:epsi_Z028}). These models not only reach lower values than our acceptance one, but also each of their parameters gets within their respective observational uncertainty listed on Table~\ref{tab:obsest} at ages of 5.170 and 5.176~Gyr \footnote{Note: these values slightly differ ($< 0.02$~Gyr) from the time of the minimum value of the epsilon squared function}.\\

\subsubsection{Mass transfer}
\begin{figure}
    \centering
    \includegraphics[width=\columnwidth]{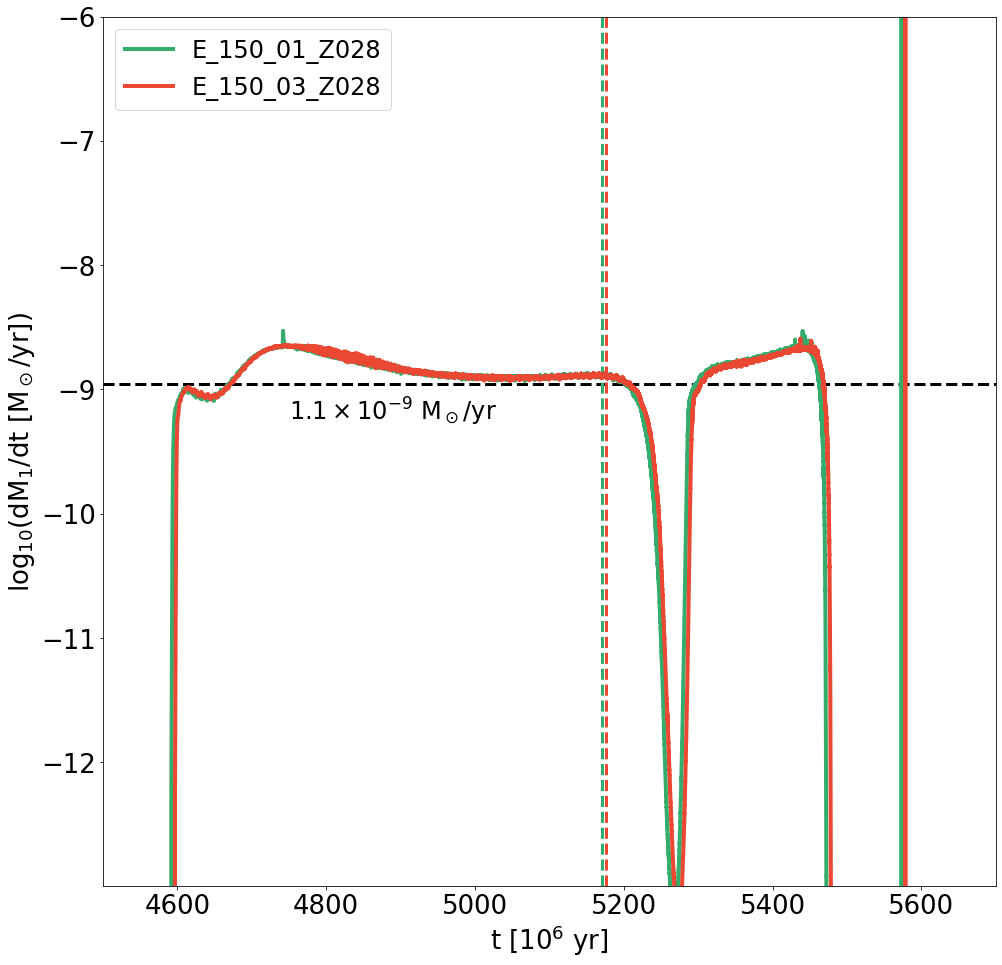}
    \caption{Donor star mass loss rate for each of the best models with the metallicity change: E\_150\_01\_Z028 and E\_150\_03\_Z028. The elements represented are the same as in Figure \ref{fig:LossRate}.\typeout{The existing estimation, made with equation 25a of \citet{Webbink83}, of $\dot{M}_{\rm d}=1.1\times10^{-9}$ is denoted with a dashed horizontal line. As for the dashed vertical lines, they represent the time when all the parameters fall within the observational uncertainties.}}
    \label{fig:LossRate_Z028}
\end{figure}

\begin{figure}
    \centering
    \includegraphics[width=\columnwidth]{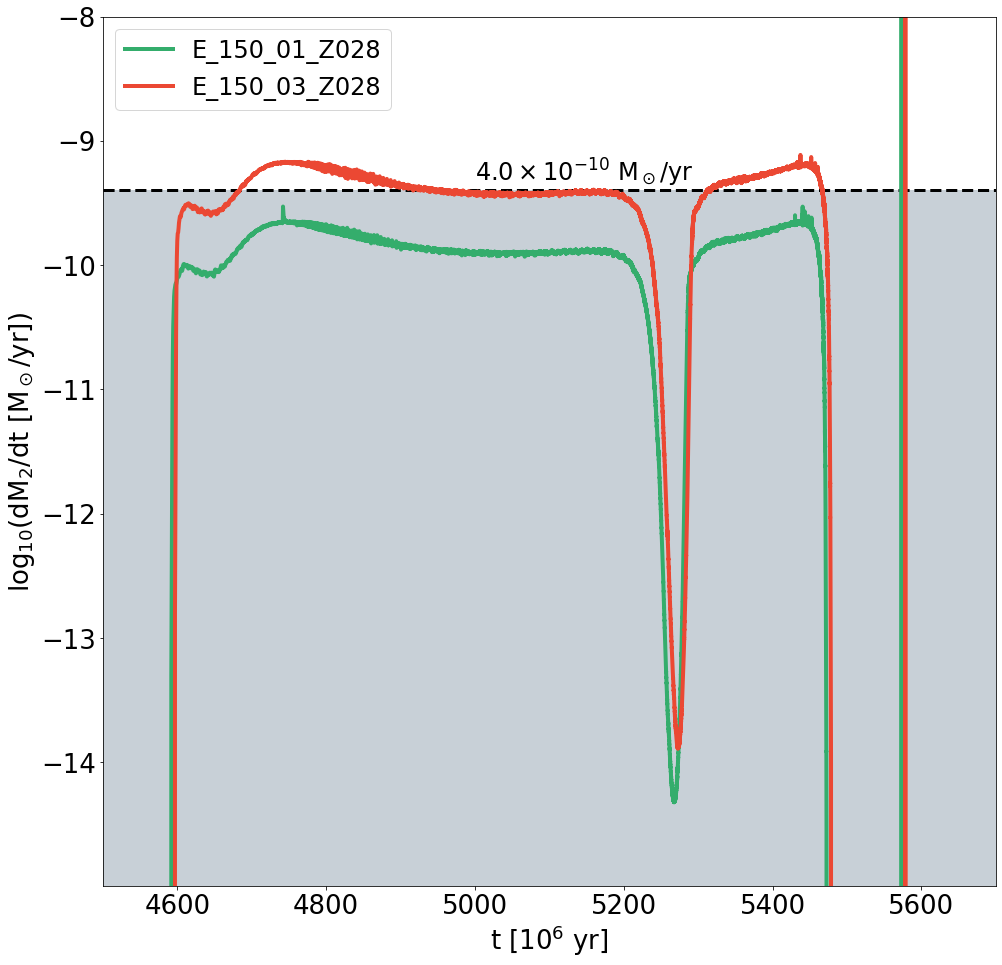}
    \caption{BH accretion rate for the two best models with double solar metallicity. The elements represented are the same as in Figure \ref{fig:AcRate}. \typeout{The qualitative form of the functions is related to the mass loss rate of the donor star by the relation $\dot{M}_{\rm 2}=-\beta\dot{M}_{\rm 1}$. This is, for a higher value for $\beta$ larger would be the accretion rate onto the compact object. The shaded area represents the zone below the likely upper limit for the mass accretion rate ($\left<\dot{M}_{\rm 2}\right>=4\times10^{-10}$) given by ZZ18.}}
    \label{fig:AcRate_Z028}
\end{figure}

A change in the metallicity of the donor star implies a change in the donor's outer opacities and, thus, in its structure as well. As for the mass transfer episode, the results can be seen in Figures~\ref{fig:LossRate_Z028}~and~\ref{fig:AcRate_Z028}. We estimate the present mass loss rate as $1.24\times10^{-9}$~M$_\odot$/yr, which is also in good accordance with the estimation given by ZZ18.\\

Considering the accretion rate, the model computed with $\beta=0.3$ seems to exceed the upper limit of $4.0\times10^{-10}$~M$_\odot$/yr at some parts of its evolution, but is below this limit near the present age. The other model, computed with $\beta = 0.1$, is still fully under the limit.\\
\begin{table}
    \centering
    \caption{Orbital period with its time derivative value and the characteristic increase time-scale evaluated on the estimated age for the system for models with Z=0.028.}
    \label{tab:dpdt_Z028}
    \resizebox{\columnwidth}{!}{
    \begin{tabular}{c|c|c|c|c}
        \hline\hline
        \textbf{Model} &  \textbf{t}$\mathbf{_{\rm min}}$ \textbf{[Gyr]} & \textbf{P}$\mathbf{_{\rm orb}}$\textbf{[d]}& $\dot{\mathbf{\rm P}}_{\mathbf{\rm orb}}$& \textbf{P}$\mathbf{_{\rm orb}}$\textbf{/}$\dot{\mathbf{\rm P}}_{\mathbf{\rm orb}}$ \textbf{[yr]}\\
        \hline
        E\_150\_01\_Z028 & 5.170 & 6.47 & $9.19\times10^{-11}$ & $1.9\times10^{8}$ \\
        E\_150\_03\_Z028 & 5.176 & 6.47 & $9.24\times10^{-11}$ & $1.9\times10^{8}$ \\
        \hline
    \end{tabular}
    }
\end{table}
\subsubsection{Orbital period and BH spin parameter}
\begin{figure}
    \centering
    \includegraphics[width=\columnwidth]{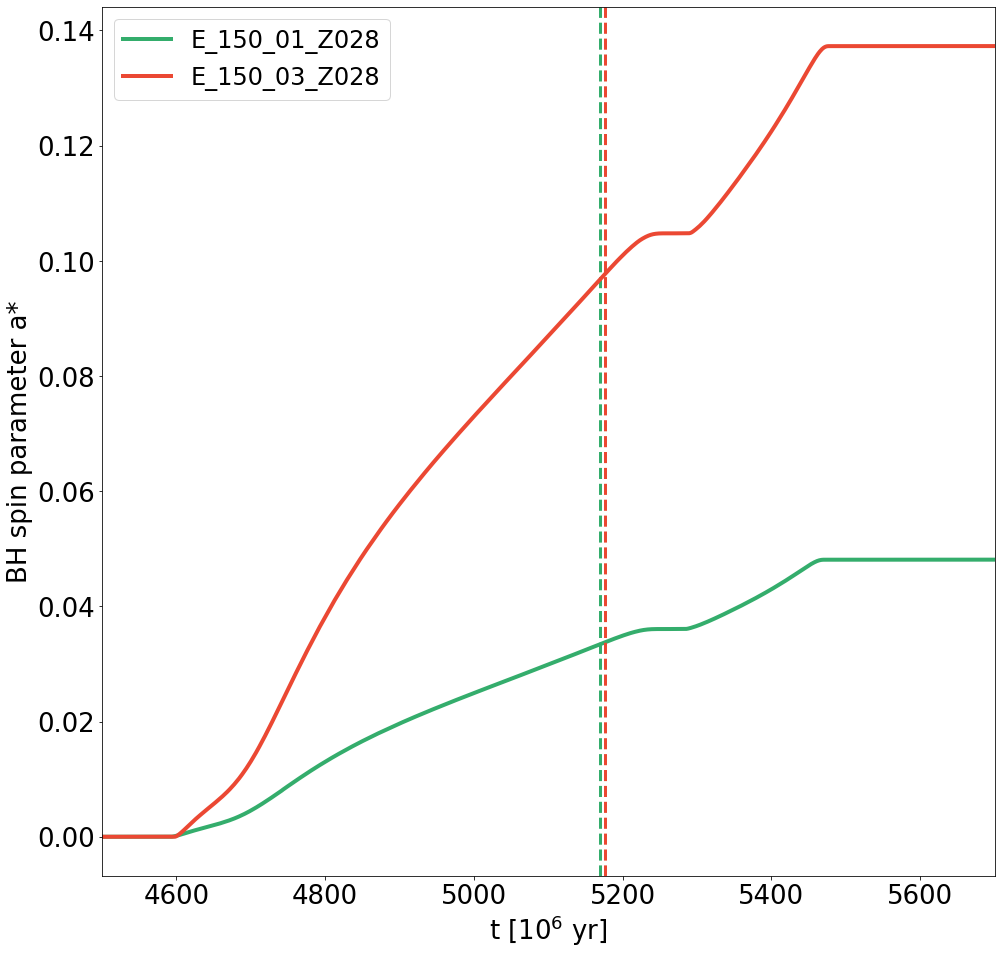}
    \caption{BH spin parameter evolution for the V404~Cyg compact object according to Equation~(\ref{BHspinparameter}). The values of the BH spin for the predicted ages for V404~Cyg (vertical dashed lines) are very similar to the two best models computed with solar composition in Table~\ref{tab:BHparam}.}
    \label{fig:BHspinparam_Z028}
\end{figure}

\begin{figure*}
    \centering
    \includegraphics[width=0.7\textwidth]{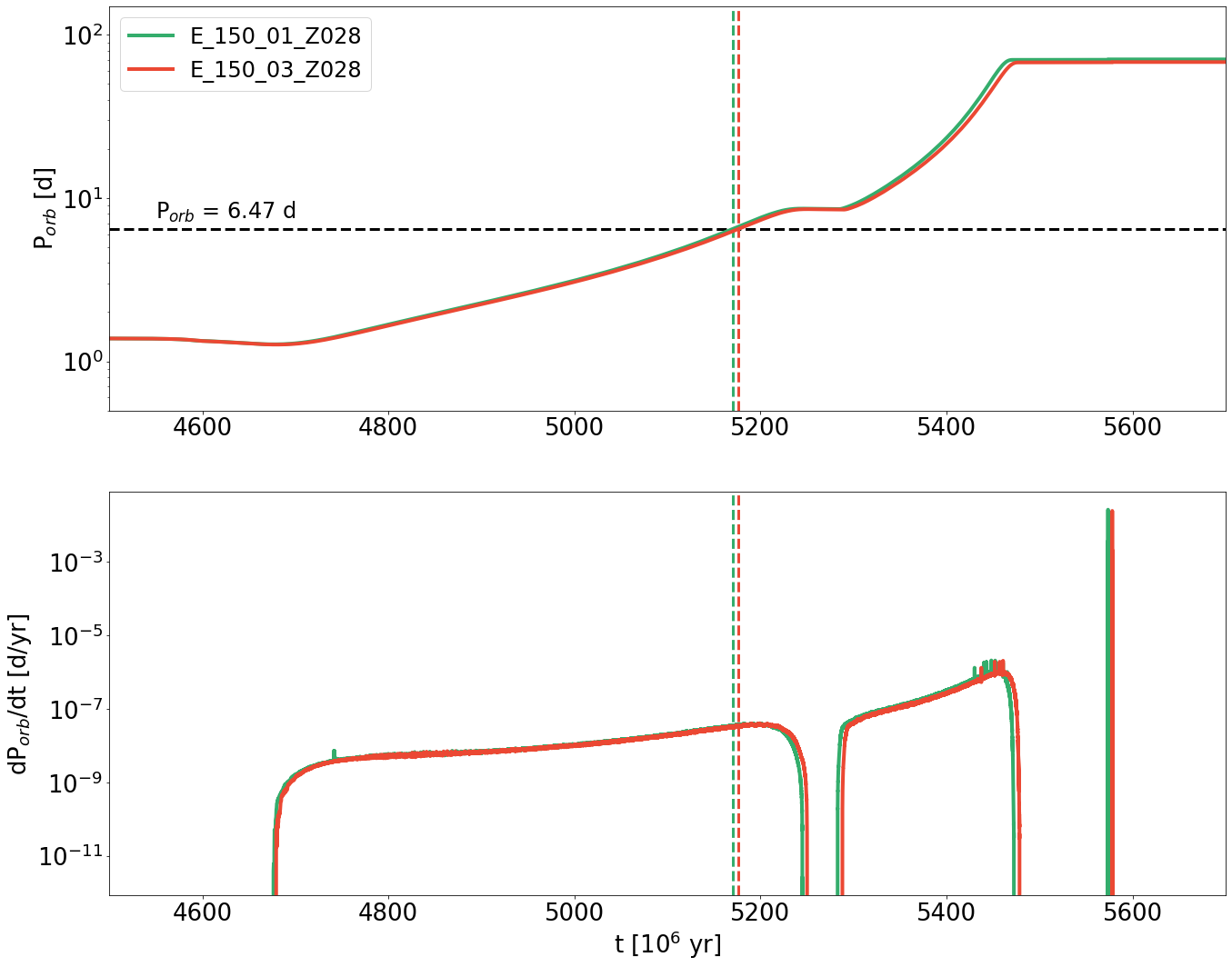}
    \caption{Orbital period and its time derivative as a function of time for models with a metallicity of Z=0.028. The represented elements are the same as the ones shown in Figure \ref{fig:DPDT} but adequate to these models.}
    \label{fig:DPDT_Z028}
\end{figure*}

As the BH accretion episode has not changed quantitatively, a significant amount from the models with lower metallicity, it is expected that the results for the BH spin parameter do not differ too much from the ones presented above. This parameter evolution for the two models that have been taken into account is shown in Figure~\ref{fig:BHspinparam_Z028}. Once again, our models do not reach the observational estimation given by~\cite{Walton2017}.\\
The orbital period evolution considering these models can be resumed in Figure~\ref{fig:DPDT_Z028} with some results given in Table~\ref{tab:dpdt_Z028}. Models with this metallicity and initial orbital period reach the observed one for V404~Cyg at the same moment when the other quantities are still within their observational uncertainties while predicting a present value of $\dot{P}_{orb} \sim 9.2\times10^{-11}$. This value deduces a characteristic increase time consistent with the one obtained by ZZ18. Our models with this metallicity deduced final orbital periods between 68 - 70 d.\\

\subsection{Donor star evolution and proposal of our bests progenitors}
\begin{figure*}
    \centering
    \includegraphics[width=0.75\textwidth]{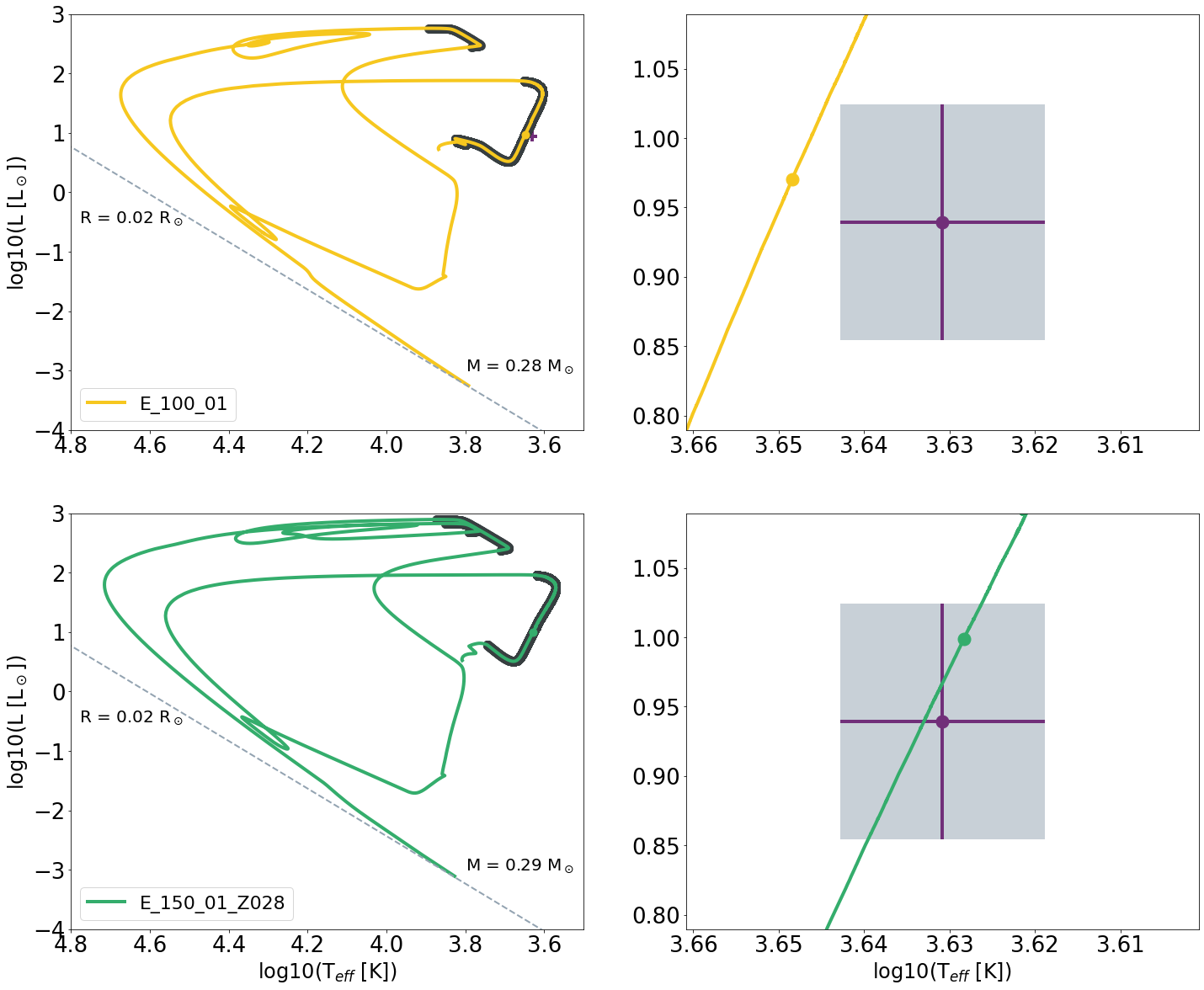}
    \caption{Evolutionary track in an HR diagram for two of our models.\textit{Top:} Model E\_100\_01 for the case of solar metallicity. \textit{Bottom:} Model E\_150\_01\_Z028 for the case of higher metallicity. The cases of the best models with higher $\beta$ value are practically overlapped with the models shown. On the left panels, the whole track is shown with a shaded area when the mass loss episode occurs. There can be seen the observationally estimated position of the donor star and the position for the age of the system predicted by our models (with circles on the evolutionary track). Once the star is getting near to becoming a white dwarf it suffers a thermonuclear flash, making it expand and lose mass again for a very short period of time. The end of these tracks are very similar, the donor becomes a low-mass white dwarf with a hydrogen-rich surface. A dashed line denoting $R = 0.02$ R$_\odot$ was included; it approximately corresponds to the asymptotic radius for the white dwarf. In the right panels, the zoom is on the nearness of this estimation, where the shaded area considers the observational error for the effective temperature and luminosity. }
    \label{fig:HR}
\end{figure*}

This work aims to model the characteristics of V404~Cyg to get a predecessor system and also to analyze its present and future evolution. With this in mind, we calculated our models up to an age of $t=14$~Gyr. This allowed us to make some estimations for the complete evolution of the system.\\
Figure~\ref{fig:HR} shows the different evolutionary tracks for the models in the Hertzsprung-Russell diagram for some of our best progenitors. Here, we can offer some remarks. As for the tracks that are calculated with solar abundances, the mass-transfer episode occurs in three parts. The first one begins on the main sequence for both models (mass transfer episode in case A; \citealt{1967ZA.....65..251K}) and ends abruptly due to the contraction of the donor when the central hydrogen is exhausted. After a very short time, the second mass transfer episode begins. This behavior cannot be appreciated in the evolutionary tracks because of the small portion of the diagram covered while the donor star is detached from its Roche lobe. Once the donor evolves blueward and gets dimmer, a hydrogen thermonuclear flash occurs making the star raise its luminosity and expand. Then, it starts the third mass transfer episode. The whole thermonuclear flash event is very fast, making the third mass transfer episode look like a Dirac's delta distribution.\\
Because the donor star is not massive enough to start the $3\alpha$ reactions\typeout{ finishing the red giant branch}, no helium burning occurs. \typeout{ Once the thermonuclear flash event has occurred, the star has lost almost 20\% of the initial superficial hydrogen abundances.}Our prediction for the final fate of the V404~Cyg donor star is to become a low-mass helium white dwarf with a mass of $0.28$~M$_\odot$ and a radius of $\sim 0.02$~R$_\odot$ with a hydrogen-rich  surface.\\
As for the present state of the donor star, observational data can be seen in Figure~\ref{fig:HR}. These data were placed on the HR diagram using the estimations of $L_{\rm d}=8.7$ L$_\odot$ and $T_{\rm eff} \sim 4200$~K and its respective error bars (see Table~\ref{tab:obsest}). Also, our theoretical models predict that the donor star is currently on the red giant branch and getting near the end of the first mass transfer episode (remaining $\sim 0.2$~Gyr).

The evolutionary tracks corresponding to a metallicity of $Z=0.028$, shown in Figure~\ref{fig:HR} (bottom panel), shares lots of characteristics in common with our previous analysis for the models with solar abundances. However, we do note that the mass transfer episode occurs in two parts, where the first one begins after core hydrogen exhaustion (case B ofmass transfer episode; \citealt{1967ZA.....65..251K}). The remnant compact object is still a helium white dwarf of mass M $\sim 0.29$ M$_\odot$. The estimated age of the system predicts that the donor star is losing mass on the main mass transfer episode and there are $\sim 0.3$ Gyr remaining for the end of it. The entirety of the mass transfer episodes takes place within $\sim 1$ Gyr.\\

\begin{figure*}
    \centering
    \includegraphics[width=0.75\textwidth]{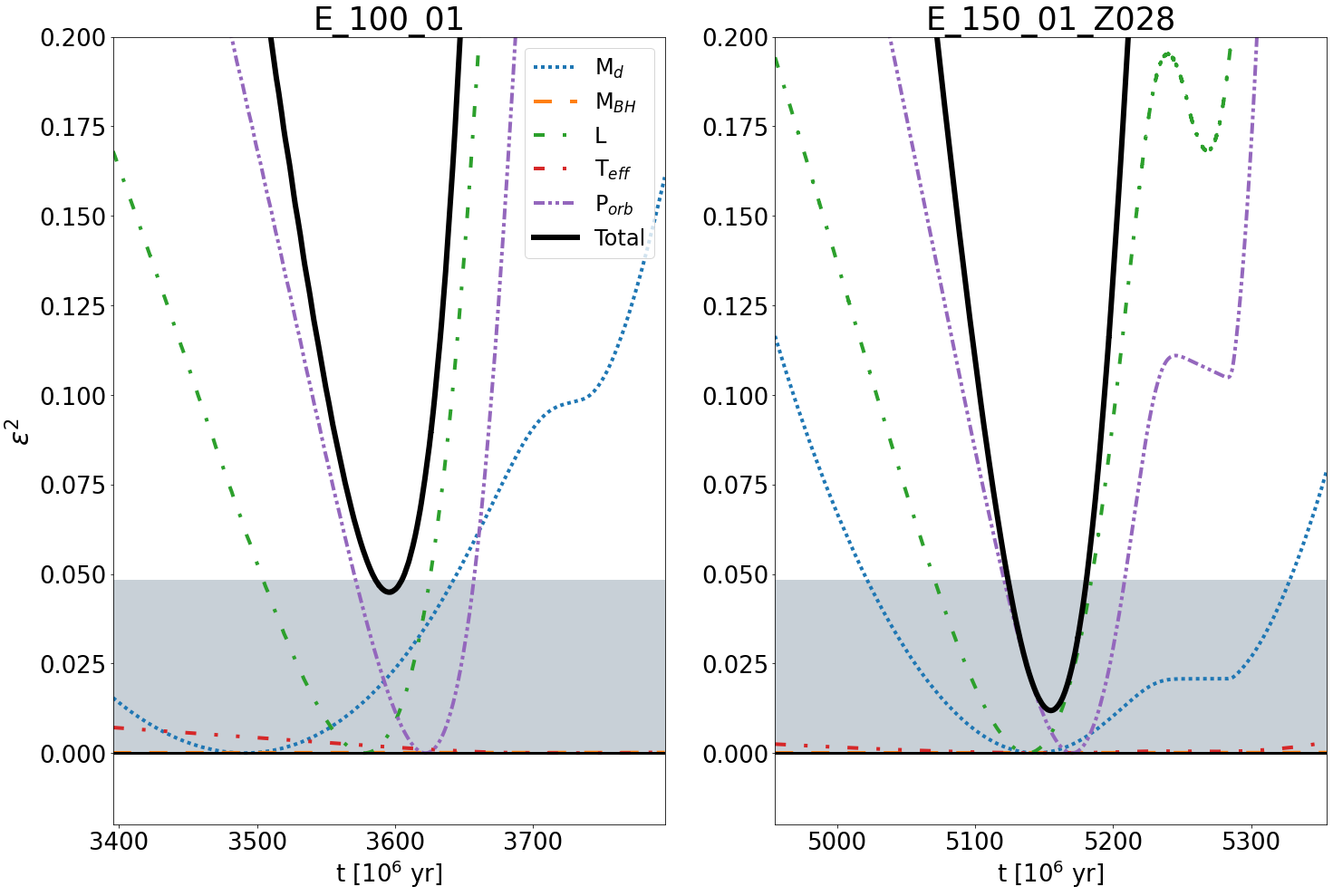}
    \caption{Epsilon squared function decomposed on the contributions of every parameter analyzed for two of our models. \textit{Left:} Model with Z=0.014. \textit{Right:} Model with Z=0.028.\typeout{The models computed with the same initial parameters except for $\beta$ have similar behavior.} The shaded area represents the acceptance zone, where the function takes values lower than $0.0485$ and the black horizontal line at $\epsilon^{2} = 0$ represents where the modeled parameter equals the observed estimation for it.}
    \label{fig:epsi-cont}
\end{figure*}

In Figure~\ref{fig:epsi-cont}, we present the contribution for every parameter considered in the construction of the epsilon squared function for one of the best models with each metallicity. For the case of solar abundances (left panel), we can see that near the minimum value, the parameters that dominate the total epsilon squared function are the system orbital period and the donor's luminosity with a great contribution of the donor's mass, while with Z=0.028 (right panel) the best models get a better estimation for this last parameter but it's behavior is still dominated by the luminosity and orbital period. The evolution of binaries is very sensitive to the variation of these parameters since the orbital evolution of the system takes a primary role in the RLOF episodes and the donor's initial mass determines the initial position on the ZAMS and the way it evolves. So, even with a thin grid on these parameters, we would have to get very specific for these initial parameters to find better models. As for the donor's luminosity, this is the observed parameter that has the largest relative error. This is due to the fact that the system is located in the bulge of our galaxy, which is a very obscured area, so we do not rely much on this quantity.\\


\section{Conclusions}\label{sec:conclusions}

Based on the calculations and analysis we performed in this work, we are in a position to propose a plausible progenitor for V404~Cyg. We considered, in the first step of calculations, solar abundances, and in the second step, a more metallic donor star. From the results given by our models and the analysis performed in the previous section, we found that two models with each metallicity considered could represent adequately the current state of V404~Cyg.  

Considering the epsilon squared ($\epsilon^2$) function (see Equation~\ref{eq:epsilon}), we selected the two best models for each metallicity:
E\_100\_01 reaching $\epsilon^2_{\rm min} = 0.0449$ and E\_100\_03 with $\epsilon^2_{\rm min} = 0.0482$ (solar abundances, Z=0.014), and E\_150\_01\_Z028 with  $\epsilon^2_{\rm min} = 0.0119$ and E\_150\_03\_Z028, with  $\epsilon^2_{\rm min} = 0.0132$ (Z=0.028). The last two, not only reached minimum values lower than the accepted one but also each quantity is{\it simultaneously} within its observational uncertainty.  

Then, we consider that the best progenitor for V404~Cyg is a system formed by a BH of $9~M_{\odot}$ together with a normal star of $1.5~M_{\odot}$, with a metal content of $Z=0.028$. The orbital period of this progenitor was $1.5$ d, and the BH accretes between $10$ and $30 \%$ of the mass lost by its companion.  

This model predicts that the donor star of this system may have a hydrogen thermonuclear flash event leading to a short mass loss episode. The remnant of the evolution for the donor star is predicted to be a low-mass helium white dwarf with a hydrogen-rich envelope of mass M~=~0.29 M$_\odot$ and radii R = 0.02 R$_\odot$.\\
Although most of the main characteristics of the V404~Cyg system are accounted for by our models, there is one specifically that is not. This is the BH spin parameter, for which we obtained values that are far below the only observational available one, presented in \citet{Walton2017}. It seems natural to consider that this discrepancy is due to the assumption that the BH is initially not rotating. Thus, our results may be interpreted as giving some evidence that in the context of close binary systems, stellar mass BHs may be born with appreciable angular momentum.

\begin{acknowledgements}
      The authors want to thank our anonymous referee for his/her comments and suggestions that helped us to improve our work. They also thank Florencia Vieyro and Federico García, whose comments were of very good use.
\end{acknowledgements}

\bibliographystyle{aa}
\bibliography{biblio}

\begin{thebibliography}{43}
\expandafter\ifx\csname natexlab\endcsname\relax\def\natexlab#1{#1}\fi

\bibitem[{{Asplund} {et~al.}(2021){Asplund}, {Amarsi}, \&
  {Grevesse}}]{2021A&A...653A.141A}
{Asplund}, M., {Amarsi}, A.~M., \& {Grevesse}, N. 2021, \aap, 653, A141

\bibitem[{{Bardeen}(1970)}]{Bardeen70}
{Bardeen}, J.~M. 1970, \nat, 226, 64

\bibitem[{{Barthelmy} {et~al.}(2015){Barthelmy}, {Chester}, {Malesani}, \&
  {Page}}]{Barthelmy2015}
{Barthelmy}, S.~D., {Chester}, M.~M., {Malesani}, D., \& {Page}, K.~L. 2015,
  GRB Coordinates Network, 17949, 1

\bibitem[{{Benvenuto} \& {De Vito}(2003)}]{Benvenuto2003}
{Benvenuto}, O.~G. \& {De Vito}, M.~A. 2003, \mnras, 342, 50

\bibitem[{{Benvenuto} {et~al.}(2012){Benvenuto}, {De Vito}, \&
  {Horvath}}]{BDVHa}
{Benvenuto}, O.~G., {De Vito}, M.~A., \& {Horvath}, J.~E. 2012, \apjl, 753, L33

\bibitem[{{Casares} \& {Charles}(1994)}]{Casares94}
{Casares}, J. \& {Charles}, P.~A. 1994, \mnras, 271, L5

\bibitem[{{Casares} {et~al.}(1992){Casares}, {Charles}, \&
  {Naylor}}]{Casares92}
{Casares}, J., {Charles}, P.~A., \& {Naylor}, T. 1992, \nat, 355, 614

\bibitem[{{Charles} {et~al.}(1989){Charles}, {Casares}, {Jones}, {Broadhurst},
  {Callanan}, {Carter}, {Hacking}, {Hassall}, {Lawrence}, {Naylor}, {Rutten},
  {Sahu}, \& {Taylor}}]{1989ESASP.296..103C}
{Charles}, P.~A., {Casares}, J., {Jones}, D.~H.~P., {et~al.} 1989, in ESA
  Special Publication, Vol.~1, Two Topics in X-Ray Astronomy, Volume 1: X Ray
  Binaries. Volume 2: AGN and the X Ray Background, ed. J.~{Hunt} \&
  B.~{Battrick}, 103

\bibitem[{{Chen} {et~al.}(1997){Chen}, {Shrader}, \& {Livio}}]{Chen97}
{Chen}, W., {Shrader}, C.~R., \& {Livio}, M. 1997, \apj, 491, 312

\bibitem[{{Cox}(2000)}]{Cox2000}
{Cox}, A.~N. 2000, {Allen's astrophysical quantities} (Springer Link)

\bibitem[{{De Vito} \& {Benvenuto}(2012)}]{DeVito2012}
{De Vito}, M.~A. \& {Benvenuto}, O.~G. 2012, \mnras, 421, 2206

\bibitem[{{Ferguson} {et~al.}(2005){Ferguson}, {Alexander}, {Allard}, {Barman},
  {Bodnarik}, {Hauschildt}, {Heffner-Wong}, \& {Tamanai}}]{Feguson2005}
{Ferguson}, J.~W., {Alexander}, D.~R., {Allard}, F., {et~al.} 2005, \apj, 623,
  585

\bibitem[{{Fukumura} {et~al.}(2021){Fukumura}, {Kazanas}, {Shrader}, {Tombesi},
  {Kalapotharakos}, \& {Behar}}]{2021ApJ...912...86F}
{Fukumura}, K., {Kazanas}, D., {Shrader}, C., {et~al.} 2021, \apj, 912, 86

\bibitem[{{Giacconi} {et~al.}(1962){Giacconi}, {Gursky}, {Paolini}, \&
  {Rossi}}]{1962PhRvL...9..439G}
{Giacconi}, R., {Gursky}, H., {Paolini}, F.~R., \& {Rossi}, B.~B. 1962, \prl,
  9, 439

\bibitem[{{Gonz{\'a}lez Hern{\'a}ndez} {et~al.}(2011){Gonz{\'a}lez
  Hern{\'a}ndez}, {Casares}, {Rebolo}, {Israelian}, {Filippenko}, \&
  {Chornock}}]{2011ApJ...738...95G}
{Gonz{\'a}lez Hern{\'a}ndez}, J.~I., {Casares}, J., {Rebolo}, R., {et~al.}
  2011, \apj, 738, 95

\bibitem[{{Iglesias} \& {Rogers}(1996)}]{Iglesias96}
{Iglesias}, C.~A. \& {Rogers}, F.~J. 1996, \apj, 464, 943

\bibitem[{{Ivanova} {et~al.}(2017){Ivanova}, {da Rocha}, {Van}, \&
  {Nandez}}]{2017ApJ...843L..30I}
{Ivanova}, N., {da Rocha}, C.~A., {Van}, K.~X., \& {Nandez}, J. L.~A. 2017,
  \apjl, 843, L30

\bibitem[{{Khargharia} {et~al.}(2010){Khargharia}, {Froning}, \&
  {Robinson}}]{Khargharia2010}
{Khargharia}, J., {Froning}, C.~S., \& {Robinson}, E.~L. 2010, \apj, 716, 1105

\bibitem[{{Kimura} {et~al.}(2016){Kimura}, {Isogai}, {Kato}, {Ueda},
  {Nakahira}, {Shidatsu}, {Enoto}, {Hori}, {Nogami}, {Littlefield}, {Ishioka},
  {Chen}, {King}, {Wen}, {Wang}, {Lehner}, {Schwamb}, {Wang}, {Zhang},
  {Alcock}, {Axelrod}, {Bianco}, {Byun}, {Chen}, {Cook}, {Kim}, {Lee},
  {Marshall}, {Pavlenko}, {Antonyuk}, {Antonyuk}, {Pit}, {Sosnovskij},
  {Babina}, {Baklanov}, {Pozanenko}, {Mazaeva}, {Schmalz}, {Reva}, {Belan},
  {Inasaridze}, {Tungalag}, {Volnova}, {Molotov}, {de Miguel}, {Kasai},
  {Stein}, {Dubovsky}, {Kiyota}, {Miller}, {Richmond}, {Goff}, {Andreev},
  {Takahashi}, {Kojiguchi}, {Sugiura}, {Takeda}, {Yamada}, {Matsumoto},
  {James}, {Pickard}, {Tordai}, {Maeda}, {Ruiz}, {Miyashita}, {Cook}, {Imada},
  \& {Uemura}}]{Kimura2016a}
{Kimura}, M., {Isogai}, K., {Kato}, T., {et~al.} 2016, \nat, 529, 54

\bibitem[{{King} \& {Lasota}(2021)}]{2021arXiv211203779K}
{King}, A. \& {Lasota}, J.-P. 2021, arXiv e-prints, arXiv:2112.03779

\bibitem[{{King} \& {Kolb}(1999)}]{King99}
{King}, A.~R. \& {Kolb}, U. 1999, \mnras, 305, 654

\bibitem[{{Kippenhahn} \& {Weigert}(1967)}]{1967ZA.....65..251K}
{Kippenhahn}, R. \& {Weigert}, A. 1967, \zap, 65, 251

\bibitem[{{Landau} \& {Lifshitz}(1971)}]{Landau71}
{Landau}, L.~D. \& {Lifshitz}, E.~M. 1971, {The classical theory of fields}
  (Pergamon Press)

\bibitem[{{Langer} {et~al.}(2020){Langer}, {Sch{\"u}rmann}, {Stoll},
  {Marchant}, {Lennon}, {Mahy}, {de Mink}, {Quast}, {Riedel}, {Sana},
  {Schneider}, {Schootemeijer}, {Wang}, {Almeida}, {Bestenlehner},
  {Bodensteiner}, {Castro}, {Clark}, {Crowther}, {Dufton}, {Evans}, {Fossati},
  {Gr{\"a}fener}, {Grassitelli}, {Grin}, {Hastings}, {Herrero}, {de Koter},
  {Menon}, {Patrick}, {Puls}, {Renzo}, {Sander}, {Schneider}, {Sen}, {Shenar},
  {Sim{\'o}n-D{\'\i}as}, {Tauris}, {Tramper}, {Vink}, \&
  {Xu}}]{2020A&A...638A..39L}
{Langer}, N., {Sch{\"u}rmann}, C., {Stoll}, K., {et~al.} 2020, \aap, 638, A39

\bibitem[{{Lewin} {et~al.}(1967){Lewin}, {Clark}, \&
  {Smith}}]{1967AJ.....72R.812L}
{Lewin}, W. H.~G., {Clark}, G.~W., \& {Smith}, W.~B. 1967, \aj, 72, 812

\bibitem[{{Makino}(1989)}]{Makino89}
{Makino}, F. 1989, \iaucirc, 4782, 1

\bibitem[{{Mart{\'\i}} {et~al.}(2016){Mart{\'\i}}, {Luque-Escamilla}, \&
  {Garc{\'\i}a-Hern{\'a}ndez}}]{Marti2016}
{Mart{\'\i}}, J., {Luque-Escamilla}, P.~L., \& {Garc{\'\i}a-Hern{\'a}ndez},
  M.~T. 2016, \aap, 586, A58

\bibitem[{{Mata S{\'a}nchez} {et~al.}(2021){Mata S{\'a}nchez}, {Rau},
  {{\'A}lvarez Hern{\'a}ndez}, {van Grunsven}, {Torres}, \&
  {Jonker}}]{2021MNRAS.506..581M}
{Mata S{\'a}nchez}, D., {Rau}, A., {{\'A}lvarez Hern{\'a}ndez}, A., {et~al.}
  2021, \mnras, 506, 581

\bibitem[{{Miko{\l}ajewska} {et~al.}(2022){Miko{\l}ajewska}, {Zdziarski},
  {Zi{\'o}{\l}kowski}, {Torres}, \& {Casares}}]{2022ApJ...930....9M}
{Miko{\l}ajewska}, J., {Zdziarski}, A.~A., {Zi{\'o}{\l}kowski}, J., {Torres},
  M. A.~P., \& {Casares}, J. 2022, \apj, 930, 9

\bibitem[{{Miller-Jones} {et~al.}(2009){Miller-Jones}, {Jonker}, {Dhawan},
  {Brisken}, {Rupen}, {Nelemans}, \& {Gallo}}]{Miller-Jones2009}
{Miller-Jones}, J.~C.~A., {Jonker}, P.~G., {Dhawan}, V., {et~al.} 2009, \apjl,
  706, L230

\bibitem[{{Motta} {et~al.}(2017){Motta}, {Kajava}, {S{\'a}nchez-Fern{\'a}ndez},
  {Giustini}, \& {Kuulkers}}]{Motta2017a}
{Motta}, S.~E., {Kajava}, J.~J.~E., {S{\'a}nchez-Fern{\'a}ndez}, C.,
  {Giustini}, M., \& {Kuulkers}, E. 2017, \mnras, 468, 981

\bibitem[{{Motta} {et~al.}(2016){Motta}, {Sanchez-Fernandez}, \&
  {Kajava}}]{Motta2016}
{Motta}, S.~E., {Sanchez-Fernandez}, C., \& {Kajava}, J. 2016, in 11th INTEGRAL
  Conference Gamma-Ray Astrophysics in Multi-Wavelength Perspective, 20

\bibitem[{{Mu{\~n}oz-Darias} {et~al.}(2016){Mu{\~n}oz-Darias}, {Casares}, {Mata
  S{\'a}nchez}, {Fender}, {Armas Padilla}, {Linares}, {Ponti}, {Charles},
  {Mooley}, \& {Rodriguez}}]{2016Natur.534...75M}
{Mu{\~n}oz-Darias}, T., {Casares}, J., {Mata S{\'a}nchez}, D., {et~al.} 2016,
  \nat, 534, 75

\bibitem[{{Podsiadlowski} {et~al.}(2003){Podsiadlowski}, {Rappaport}, \&
  {Han}}]{Podsi2003}
{Podsiadlowski}, P., {Rappaport}, S., \& {Han}, Z. 2003, \mnras, 341, 385
  (PRH03)

\bibitem[{{Rappaport} {et~al.}(1982){Rappaport}, {Joss}, \&
  {Webbink}}]{Rappaport82}
{Rappaport}, S., {Joss}, P.~C., \& {Webbink}, R.~F. 1982, \apj, 254, 616

\bibitem[{{Rappaport} {et~al.}(1983){Rappaport}, {Verbunt}, \&
  {Joss}}]{Rappaport83}
{Rappaport}, S., {Verbunt}, F., \& {Joss}, P.~C. 1983, \apj, 275, 713

\bibitem[{{Verbunt} \& {Zwaan}(1981)}]{Verbunt81}
{Verbunt}, F. \& {Zwaan}, C. 1981, \aap, 100, L7

\bibitem[{{Wagner} {et~al.}(1989){Wagner}, {Kreidl}, {Howell}, {Collins}, \&
  {Starrfield}}]{Wagner89}
{Wagner}, R.~M., {Kreidl}, T.~J., {Howell}, S.~B., {Collins}, G.~W., \&
  {Starrfield}, S. 1989, \iaucirc, 4797, 1

\bibitem[{{Walton} {et~al.}(2017){Walton}, {Mooley}, {King}, {Tomsick},
  {Miller}, {Dauser}, {Garc{\'\i}a}, {Bachetti}, {Brightman}, {Fabian},
  {Forster}, {F{\"u}rst}, {Gandhi}, {Grefenstette}, {Harrison}, {Madsen},
  {Meier}, {Middleton}, {Natalucci}, {Rahoui}, {Rana}, \& {Stern}}]{Walton2017}
{Walton}, D.~J., {Mooley}, K., {King}, A.~L., {et~al.} 2017, \apj, 839, 110

\bibitem[{{Webbink} {et~al.}(1983){Webbink}, {Rappaport}, \&
  {Savonije}}]{Webbink83}
{Webbink}, R.~F., {Rappaport}, S., \& {Savonije}, G.~J. 1983, \apj, 270, 678

\bibitem[{{You} {et~al.}(2023){You}, {Dong}, {Yan}, {Liu}, {Tuo}, {Yao}, \&
  {Cao}}]{2023ApJ...945...65Y}
{You}, B., {Dong}, Y., {Yan}, Z., {et~al.} 2023, \apj, 945, 65

\bibitem[{{Zi{\'o}{\l}kowski} \& {Zdziarski}(2018)}]{Ziolkowski2018}
{Zi{\'o}{\l}kowski}, J. \& {Zdziarski}, A.~A. 2018, \mnras, 480, 1580 (ZZ18)

\bibitem[{{{\.Z}ycki} {et~al.}(1999){{\.Z}ycki}, {Done}, \& {Smith}}]{Zycki99}
{{\.Z}ycki}, P.~T., {Done}, C., \& {Smith}, D.~A. 1999, \mnras, 309, 561

\end{thebibliography}
\end{document}